\documentclass[a4paper, 11pt]{article}
\usepackage{pos}
\usepackage{comment}

\title{Early charmless \textit{B} physics at Belle II}
 \ShortTitle{Early charmless \textit{B} physics at Belle II}

\author*[1]{Eldar Ganiev}
\note{On behalf of the Belle II collaboration}

\affiliation{University and INFN, Trieste, Italy}

\emailAdd{eldar.ganiev@ts.infn.it}

\abstract{We report on the first measurements of branching fractions, CP-violating charge asymmetries, and longitudinal polarization fractions in charmless $B$ decays at the Belle II experiment. We use a sample of electron-positron collisions collected in 2019 and 2020 at the $\Upsilon(4S)$ resonance that corresponds to 34.6 fb$^{-1}$ of integrated luminosity. The results are compatible with the known values, which indicates a good understanding of detector performance.}

\FullConference{
  40th International Conference on High Energy physics - ICHEP2020\\
  July 28 - August 6, 2020\\
  Prague, Czech Republic (virtual meeting)
}

\begin{document}
\maketitle

\section{Introduction}

The physics of charmless $B$ decays is an essential portion of the Belle II program~\cite{kou}. The expected large yields will enable significant advancements in the understanding of quark dynamics, including an improved determination of the CKM phase $\alpha/\phi_2$, a conclusive test of the $K\pi$ isospin sum-rule~\cite{gronau}, a thorough investigation of CP-violating asymmetries localized in the phase space of three-body $B$ decays, and measurements of the decay-time-dependent CP violation in almost pure penguin $b\to q\overline{q}s$ channels, such as $B^0\to\phi K^{0}_\textrm{S}$ and $B^0\to\eta^{\prime} K^{0}_\textrm{S}$ decays.

Belle II is a magnetic spectrometer, designed to reconstruct the products of electron-positron collisions produced by the SuperKEKB asymmetric-energy collider, located at the KEK laboratory, Japan. The Belle II detector started its collision operations on March 11, 2019. The sample of electron-positron collisions used in this work corresponds to an integrated luminosity of 34.6 fb$^{-1}$~\cite{lumi} and was collected at the $\Upsilon(4S)$ resonance as of May 14, 2020. We aim to perform first measurements in charmless $B$ decay modes, which offer ideal benchmarks to test tracking, reconstruction of neutral particles, particle identification, vertexing, and advanced analysis techniques capabilities. 

We focus on decays with branching fractions of 10$^{-6}$, or larger, into final states sufficiently simple to obtain visible signals in the analyzed data set with a relatively straightforward reconstruction. The target decay modes are two-body $B^0\to K^+\pi^-$, $B^+\to K^+\pi^0$, $B^+\to K^{0}_\textrm{S}\pi^+$, $B^0\to K^{0}_\textrm{S}\pi^0$, $B^0\to \pi^+\pi^-$, $B^+\to \pi^+\pi^0$ decays; three-body $B^+\to K^+K^-K^+$, $B^+\to K^+\pi^-\pi^+$ decays, and quasi two-body $B^+\to\phi(1020) K^+$, $B^0\to\phi(1020) K^{0}_\textrm{S}$, $B^+\to \phi(1020) K^{*}(892)^{+}$, $B^0\to\phi(1020) K^{*}(892)^0$ decays. In what follows, charge-conjugate modes are implied, and $K^{*+,0}$ and $\phi$ indicate the $K^{*}(892)^{+,0}$ and $\phi(1020)$ mesons, except when otherwise stated.

The principal challenge is to overcome the initial $\lesssim$ 10$^{-5}$ signal-to-background ratio with a selection sufficiently discriminating to isolate an abundant signal. The dominant backgrounds arise from random combinations of particles produced in $continuum$ $e^+e^-\to q\overline{q}$ ($q=u,d,s,c$) events. We use two variables known to be strongly discriminating between signal and continuum: beam-energy-constrained mass and energy difference, defined as $M_{\rm bc} \equiv \sqrt{s/(4c^4) - (p^{*}_B/c)^2}$ and $\Delta E \equiv E^{*}_{B} - \sqrt{s}/2$, where $\sqrt{s}/2$ is the half of the collision energy, $E^{*}_{B}$ and $p^{*}_{B}$ are the reconstructed energy and momentum of $B$ meson candidates, all in the $\Upsilon(4S)$ frame. For further discrimination, we use a binary boosted decision-tree classifier that makes a non-linear combination of about 30~kinematic, decay time, and event-shape variables. We train the classifier to identify statistically significant signal and background features using unbiased simulated samples.

\section{Two- and three-body decays}

We search for the decays $B^0\to K^+\pi^-$, $B^+\to K^+\pi^0$, $B^+\to K^{0}_\textrm{S}\pi^+$, $B^0\to K^{0}_\textrm{S}\pi^0$, \mbox{$B^0\to \pi^+\pi^-$}, $B^+\to \pi^+\pi^0$, $B^+\to K^+K^-K^+$, and $B^+\to K^+\pi^-\pi^+$~\cite{charmless1}. We form final-state particle candidates by applying baseline selection criteria and then combine candidates in kinematic fits consistent with the topologies of the desired decays to reconstruct $B$ candidates. \newline
\indent In the simulated samples, for each channel, we simultaneously vary the selection criteria on continuum-suppression output and charged-particle identification information to maximize S/$\sqrt{\textrm{S+B}}$, where S and B are signal and background yields, respectively, estimated in the signal region. The $\pi^0$ selection is optimized using the $B^+\to\overline{D}^0(\to K^+\pi^-\pi^0)\pi^+$ control channel reconstructed in simulated and collision data. After applying the optimized selection, we restrict samples to one candidate per event, choosing it randomly. \newline
\indent Some channels show a non-negligible fraction of $\textit{self-cross-feed}$, i.e. misreconstructed signal candidates formed by misidentified (swapped mass assignments) signal particles or combinations of signal and non-signal particles. We include the self-cross-feed component in our fit model by fixing its proportions to the expectations from simulation. \newline
\indent For the three-body decays, simulation is also used to identify and suppress contamination from peaking backgrounds. We exclude the two-body invariant mass ranges corresponding to $D^0$, $\eta_c$, and $\chi_{c1}$ decays for $B^+\to K^+K^-K^+$ and $D^0$, $\eta_c$, $\chi_{c1}$, $J/\psi$, and $\psi(2S)$ decays for $B^+\to K^+\pi^-\pi^+$ decays. In addition, we veto the genuine charmless $B^+\to K^*(892)\pi^+$ subcomponent to allow for a consistent comparison of $B^+\to K^+\pi^-\pi^+$ branching fraction with the known value~\cite{pdg}. 

\begin{figure}[!htb]
\begin{center}
\includegraphics[width=5.5cm]{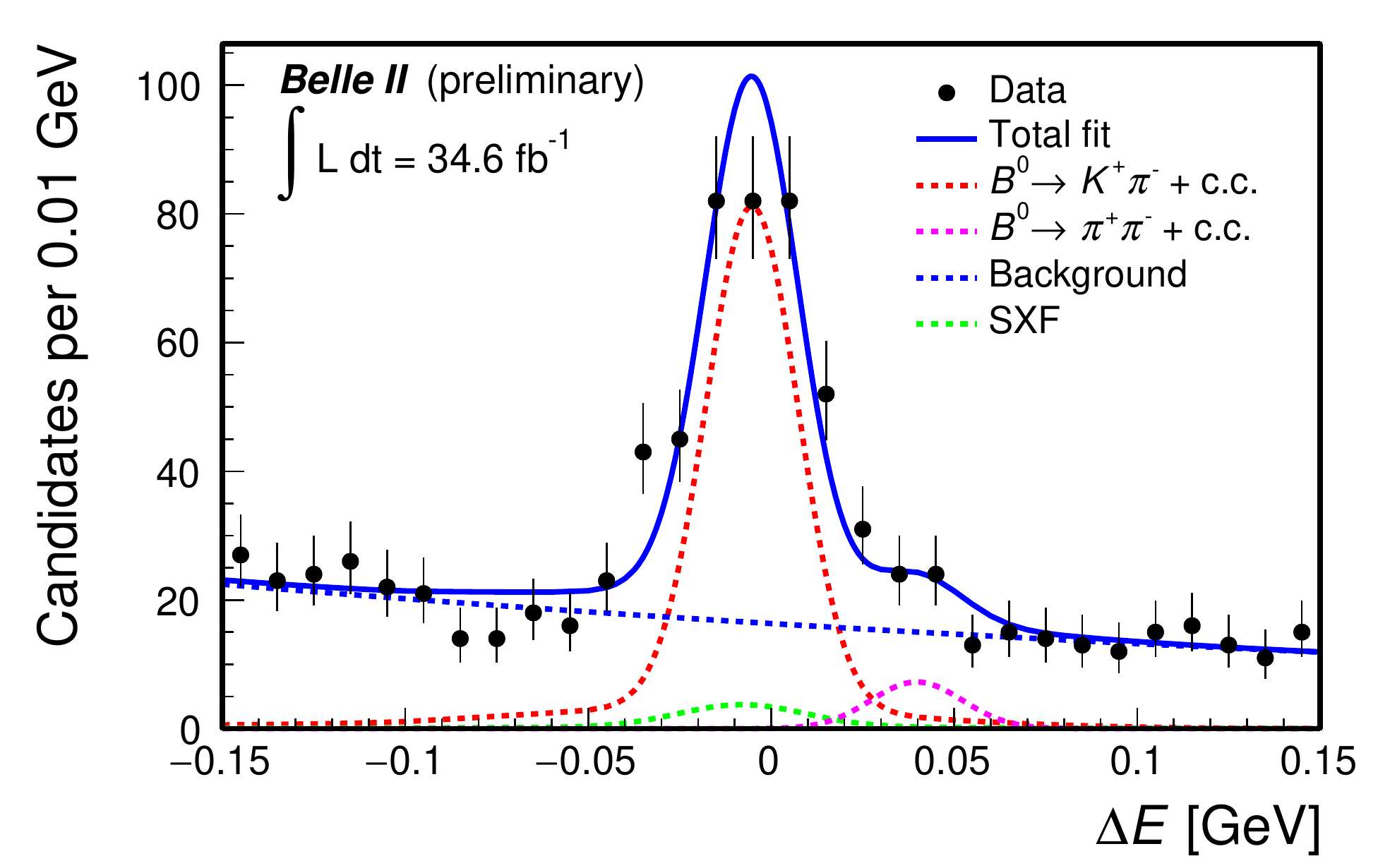}
\includegraphics[width=5.5cm]{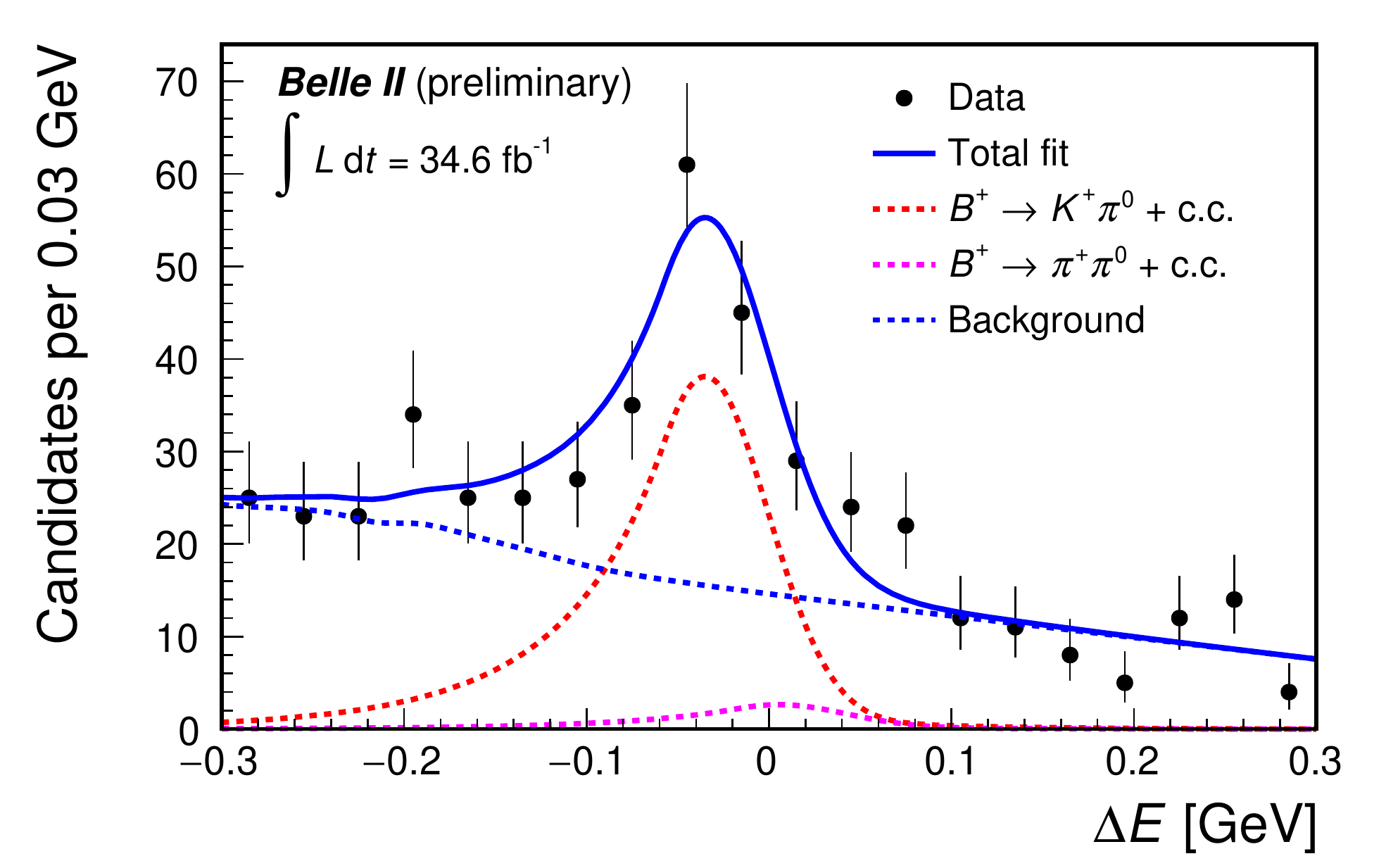}
\includegraphics[width=5.5cm]{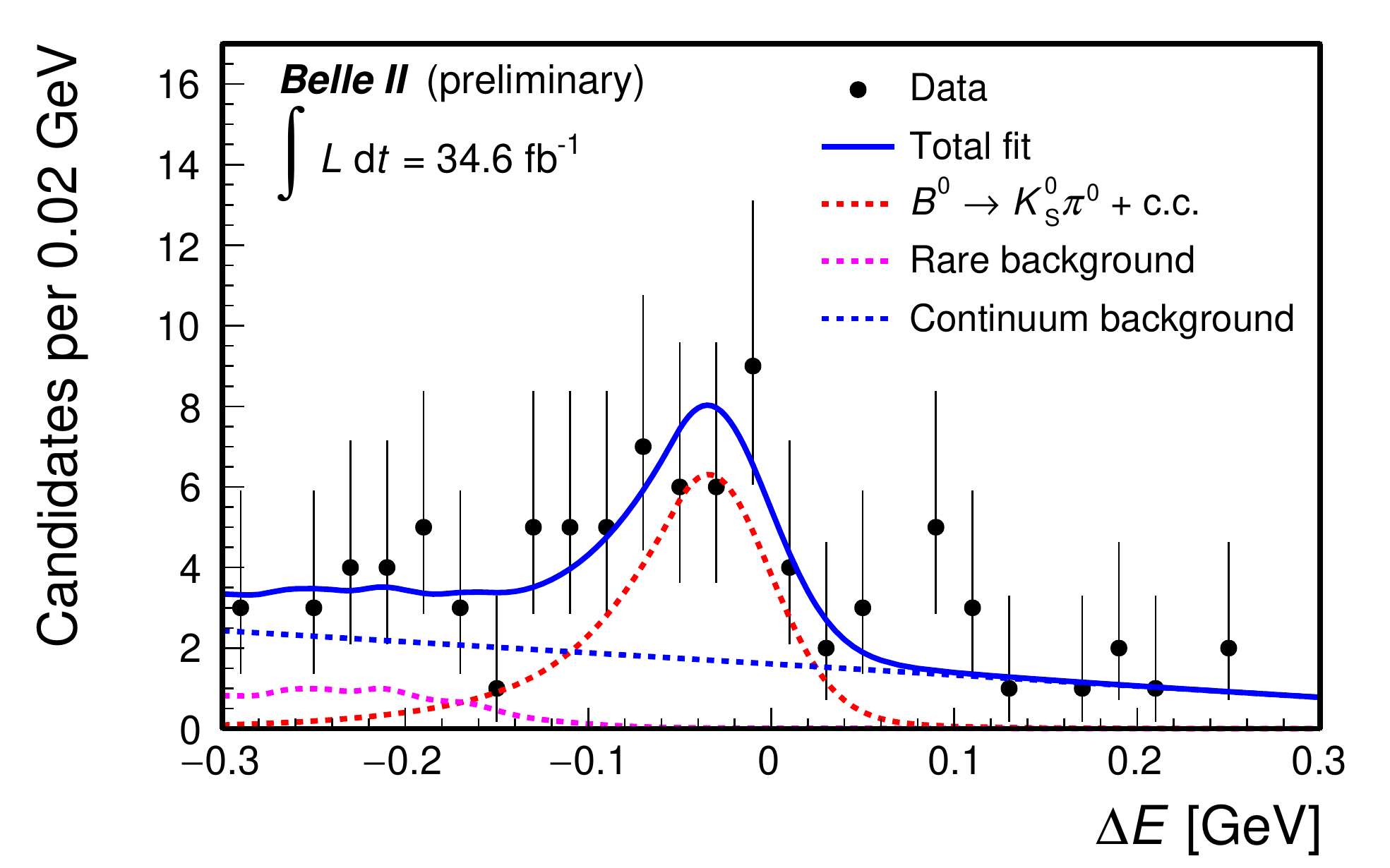}
\includegraphics[width=5.5cm]{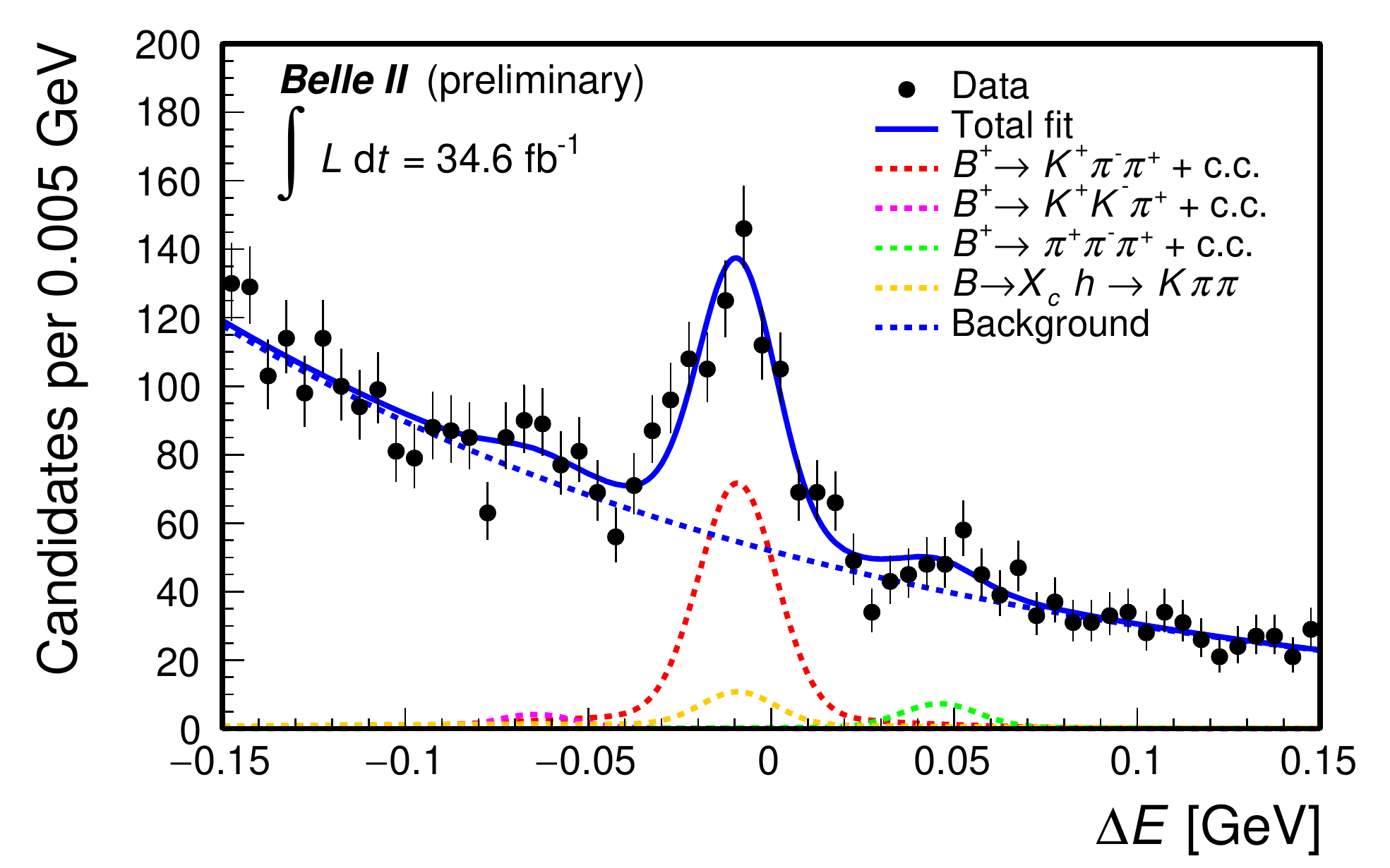}
\end{center}
 \caption{Distributions of $\Delta E$ for (top left) $B^{0}\to K^{+}\pi^{-}$, (top right) $B^{+}\to K^{+}\pi^{0}$, (bottom left) $B^{0}\to K^{0}_\textrm{S}\pi^{0}$, and (bottom right) $B^{+}\to K^{+}\pi^{-}\pi^{+}$ candidates reconstructed in 2019--2020 Belle II data. The "SXF" label indicates self-cross-feed. The projections of unbinned maximum likelihood fits are overlaid.}
 \label{fig:Kpi}
\end{figure}

We determine signal yields from maximum likelihood fits of the unbinned $\Delta E$ distributions of candidates restricted to the $M_\textrm{bc}$ > 5.27 GeV/$c^2$ and |$\Delta E$| < 0.15 GeV region. Fit models are obtained empirically from simulation, with the only additional flexibility of a shift of the signal-peak position, which is determined in data. Examples of the $\Delta E$ distributions with fit projections overlaid are shown in Fig.~\ref{fig:Kpi}.

We determine each branching fraction as $\mathcal{B} = N/(2\varepsilon N_{B\overline{B}})$, where $N$ is the signal yield obtained from the fit, $\varepsilon$ is the reconstruction and selection efficiency, and $N_{B\overline{B}}$ is the number of produced $B\overline{B}$ pairs, corresponding to 19.7 million for $B^+B^-$ and 18.7 million for $B^0\overline{B}^0$ pairs. The selection efficiencies are determined from simulation. For those efficiencies, where simulation may not accurately model data, we perform dedicated checks on control samples and assess systematic uncertainties. We obtain the number of $B\overline{B}$ pairs from the measured integrated luminosity, the $e^+e^-\to\Upsilon(4S)$ cross section (assuming that the $\Upsilon(4S)$ decays exclusively to $B\overline{B}$~pairs), and the $\Upsilon(4S)\to B^0\overline{B}^0$ branching fraction \cite{bphys}. For the branching fraction measurement of $B^+\to K^0\pi^+$ and $B^0\to K^0\pi^0$, we consider a 0.5 factor to account for the $K^0\to K^0_{\rm S}$ probability.

We also measure CP-violating asymmetries $\mathcal{A}_\textrm{CP} = \mathcal{A} - \mathcal{A}_\textrm{det}$, where $\mathcal{A}$ is the observed charge-specific signal-yield asymmetry, and $\mathcal{A}_\textrm{det}$ is the instrumental asymmetry due to differences in interaction or reconstruction probabilities between opposite-charge hadrons.  We determine $\mathcal{A}$ from a simultaneous non-extended likelihood fit of the unbinned $\Delta E$ distributions of bottom and antibottom candidates decaying in flavor-specific final states. We evaluate the instrumental asymmetries $\mathcal{A}_\textrm{det}(K^+\pi^-)$ = $-0.010 \pm 0.003$ and $\mathcal{A}_\textrm{det}(K^0_\textrm{S}\pi^+)$ = $-0.007 \pm 0.022$ by measuring the charge-asymmetry in abundant samples of $D^0\to K^-\pi^+$ and $D^+\to K^0_{\rm S}\pi^+$ decays, respectively, assuming no CP violation. Moreover, we estimate the instrumental asymmetry related to charged kaon reconstruction alone $\mathcal{A}_\textrm{det}(K^+)$ = $-0.015 \pm 0.022$ by combining all inputs in the relationship $\mathcal{A}_\textrm{det}(K^+)=\mathcal{A}_\textrm{det}(K^+\pi^-)-\mathcal{A}_\textrm{det}(K^0_\textrm{S}\pi^+)+\mathcal{A}_\textrm{det}(K^0_\textrm{S})$, where $\mathcal{A}_\textrm{det}(K^0_\textrm{S})$ estimated following Ref.~\cite{lhcb}.

Finally, we assess the main systematic effects, such as coming from tracking, $\pi^0$ reconstruction, particle identification, and shape modelling. The measured branching fractions and CP-violating asymmetries are summarized in Table~\ref{tab:res1}. The results are compatible with values measured previously.

\section{Quasi two-body decays}

We search for the decays $B^+\to\phi K^+$, $B^+\to \phi K^{*+}$, $B^0\to\phi K^{0}_\textrm{S}$, and $B^0\to\phi K^{*0}$, followed by the $\phi\to K^+K^-$, $K^{*0}\to K^+\pi^-$, $K^{*+}\to K^{0}_\textrm{S}\pi^+$, and $K^{0}_\textrm{S}\to\pi^+\pi^-$ decays \cite{gaz}. We first reconstruct charged pion and kaon candidates by applying simple track-quality and particle-identification selections. We combine them into intermediate-resonance candidates, which are required to meet invariant-mass conditions. Then, reconstructed particles are combined into $B$ meson candidates that are required to satisfy $M_\textrm{bc}$~>~5.25~GeV/$c^2$ and |$\Delta E$|~<~0.2~GeV. For each decay mode we accept one signal candidate per event retaining the $B$ candidate with the highest vertex fit probability. The fraction of self-cross-feed candidates is fixed to the expectations from simulation. To reduce remaining background, we first apply high-efficiency continuum suppression selection. Then, for each individual final state, we train and optimize a multivariate boosted decision tree classifier.

\begin{figure}[!htb]
\begin{center}
\includegraphics[width=10cm]{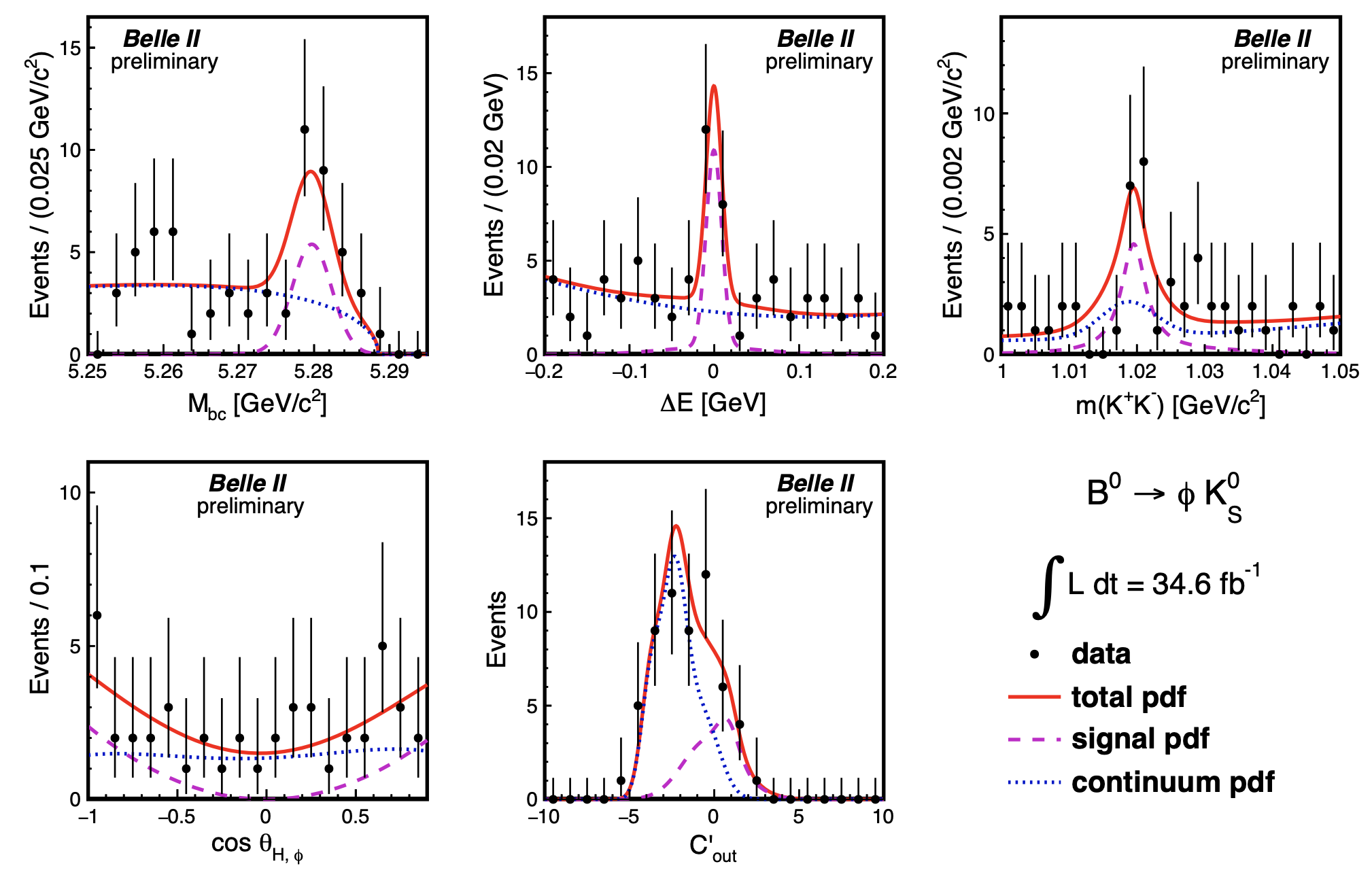}
\end{center}
 \caption{Distributions of discriminating observables of the $B^0\to\phi K^{0}_\textrm{S}$ candidates reconstructed in 2019--2020 Belle II data. The projection of unbinned maximum likelihood fit is overlaid.}
 \label{fig:phiK}
\end{figure}

\begin{figure}[!htb]
\begin{center}
\includegraphics[width=14cm]{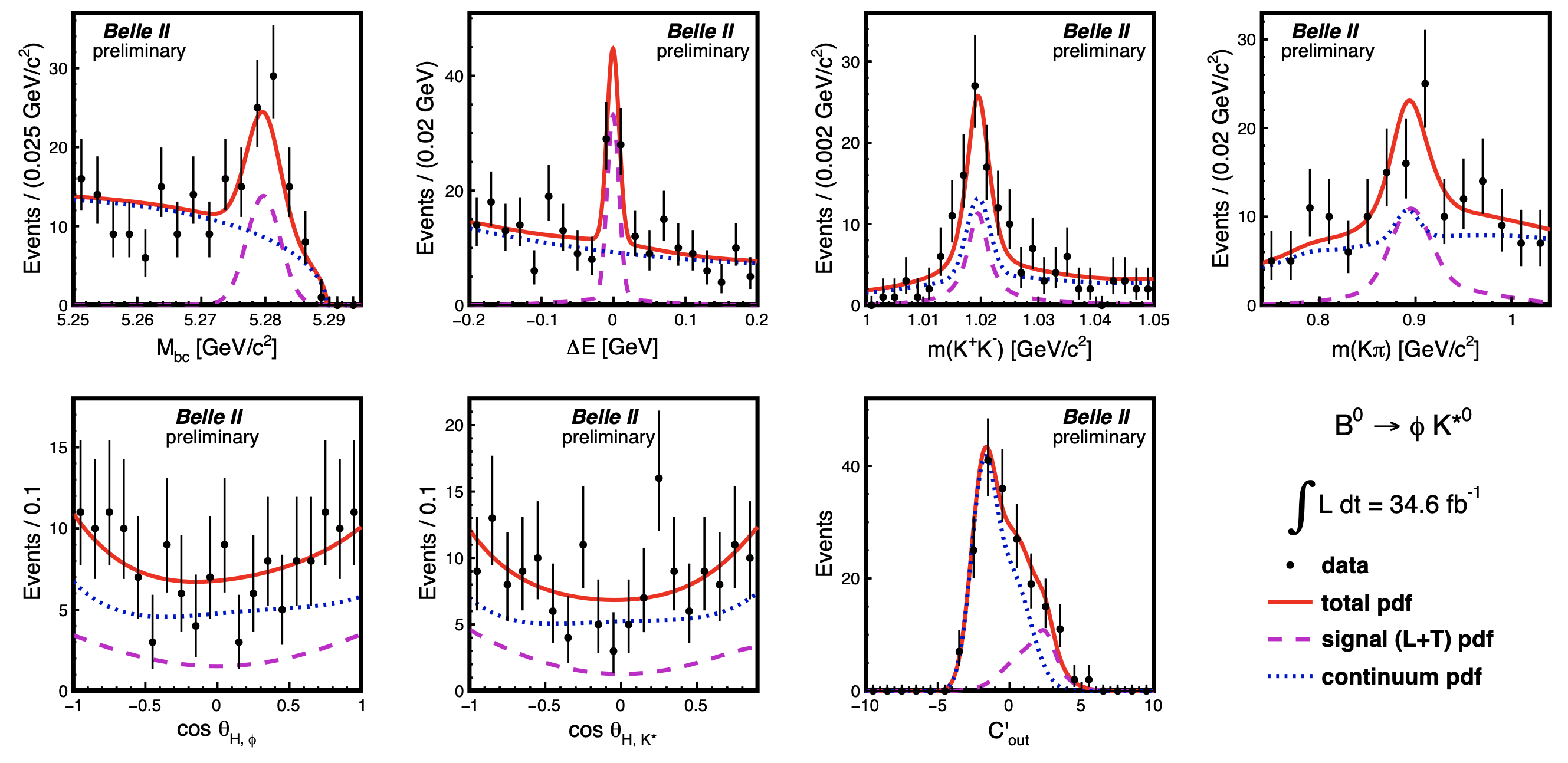}
\end{center}
 \caption{Distributions of discriminating observables of the $B^0\to\phi K^{*0}$ candidates reconstructed in 2019--2020 Belle II data. The projection of unbinned maximum likelihood fit is overlaid.}
 \label{fig:phiKst}
\end{figure}

We determine signal yields with an unbinned extended maximum likelihood fit. For each component, we compose the single event likelihood as the product of one-dimensional probability density functions for each of the observables. The observables in the fit are $M_\textrm{bc}$, $\Delta E$, $C^{\prime}_\textrm{out}$ (output of the continuum suppression classifier), $m(K^+K^-)$ (invariant mass of the $\phi$ candidate), cos$\theta_{H,\phi}$ (cosine of the helicity angle of the $\phi$ candidate), $m(K^+\pi)$ (invariant mass of the $K^*$ candidate), cos$\theta_{H,K^*}$ (cosine of the helicity angle of the $K^*$ candidate). The last two observables are relevant only for the $B\to \phi K^{*}$ modes. Fit models are obtained from simulation. Figures~\ref{fig:phiK} and~\ref{fig:phiKst} show examples of data distributions with fit projections overlaid.

Finally, we calculate branching fractions and the fraction of longitudinally polarized events $f_L$~=~$\frac{N_L/\varepsilon_L}{N_L/\varepsilon_L+N_T/\varepsilon_T}$ (if applicable), where $N_{L}$ and $N_T$ correspond to the number of longitudinally and transversely polarized signal candidates, respectively, and $\varepsilon_{L,T}$ is the corresponding selection efficiency determined from simulation. The longitudinal (transverse) polarization state corresponds to both $\phi$ and $K^*$ having zero (opposite) spin projections along the decay axis. 

We evaluate systematic uncertainties, such as those coming from tracking, particle identification, and shape modelling. The final results, summarized in Table~\ref{tab:res1}, agree with known values.

\begin{table}[h]
\centering
\footnotesize
\caption{Summary of branching fraction, CP-violating asymmetry, and longitudinal polarization fraction measurements, where the first contribution to the uncertainty is statistical, the second is systematic.}
\begin{tabular}{l c c c}
\hline\hline
 Decay mode & Branching fraction $\times~10^{-6}$ & CP-violating asymmetry & Longitudinal fraction \\
 \hline
 $B^{0}\to K^{+}\pi^{-}$ &  $18.9 \pm 1.4 \pm 1.0$ & $0.030 \pm 0.064 \pm 0.008$ & - \\[0.06cm]
 $B^+ \to K^+\pi^0$ & $12.7 ^{+2.2}_{-2.1}\pm 1.1$ & $0.052 ^{+0.121}_{-0.119}\pm 0.022$ & -  \\[0.06cm]
 $B^+ \to K^0\pi^+$ & $21.8 ^{+3.3}_{-3.0} \pm 2.9$ & $-0.072 ^{+0.109}_{-0.114} \pm 0.024$ & -  \\[0.06cm]
 $B^0 \to K^0\pi^0$ & $10.9^{+2.9}_{-2.6} \pm 1.6$ & -  & -  \\[0.06cm]
 $B^0 \to \pi^+\pi^-$ & $5.6 ^{+1.0}_{-0.9} \pm 0.3$ & -  & -  \\[0.06cm]
 $B^+ \to \pi^+\pi^0$ & $5.7 \pm 2.3\pm 0.5$ & $-0.268 ^{+0.249}_{-0.322}\pm 0.123$ & -  \\[0.06cm]
 $B^+ \to K^+K^-K^+$  & $32.0 \pm 2.2 \pm 1.4$ & $-0.049 \pm 0.063 \pm 0.022 $ & -  \\[0.06cm]
 $B^+ \to K^+\pi^-\pi^+$ & $48.0 \pm 3.8\pm 3.3$ & $-0.063 \pm 0.081 \pm 0.023$ & -  \\[0.06cm]
 $B^+\to\phi K^+$ & $6.7 \pm 1.1 \pm 0.5$& - & -  \\[0.06cm]
 $B^0\to\phi K^{0}$ & $5.9 \pm 1.8 \pm 0.7$& - & -  \\[0.06cm] 
 $B^+\to \phi K^{*+}$ & $21.7 \pm 4.6 \pm 1.9$& - & $0.58 \pm 0.23 \pm 0.02$ \\[0.06cm]
 $B^0\to \phi K^{*0}$ & $11.0 \pm 2.1 \pm 1.1$& - & $0.57 \pm 0.20 \pm 0.04$ \\[0.06cm]
\hline\hline
\end{tabular} 

\label{tab:res1}
\end{table}

\section{Summary}

We report on first measurements of branching fractions, CP-violating charge asymmetries, and longitudinal polarization fractions in charmless $B$ decays at Belle II. We use a sample of 2019 and 2020 data corresponding to 34.6 fb$^{-1}$ of integrated luminosity. From the maximum likelihood fit, we determine signal yields for the decay modes $B^0\to K^+\pi^-$, $B^+\to K^+\pi^0$, $B^+\to K^{0}_\textrm{S}\pi^+$, $B^0\to K^{0}_\textrm{S}\pi^0$, $B^0\to \pi^+\pi^-$, $B^+\to \pi^+\pi^0$, $B^+\to K^+K^-K^+$, $B^+\to K^+\pi^-\pi^+$, $B^+\to\phi K^+$, $B^+\to \phi K^{*+}$, $B^0\to\phi K^{0}_\textrm{S}$, and $B^0\to\phi K^{*0}$. The results agree with known values and show a good understanding of detector performance offering a reliable basis to assess projections for future reach.

\end{document}